\documentclass{article}
\usepackage{epsf,amsfonts,amssymb}
\usepackage{amssymb}
\usepackage{amsmath}

\begin{document}

\title{The properties of quasispecies dynamics in molecular evolution }
\author{V. V. Gafiychuk \thanks{%
Institute for Applied problem of Mechanics and Mathematics at the National
Academy of Sciences of Ukraine} and A. K. Prykarpatsky\thanks{%
Dept. of Applied Mathematics at the AGH University of Science and
Technology, Cracow, Poland and Institute for Applied Problems of Mechanics
and Mathematics at the National Academy of Sciences{}
Mailling~address:~Institute for Applied Problems of Mechanics and
Mathematics at the National Academy of Sciences{}, Naukova
St,~3-b,~~Lviv~79601,~Ukraine}}
\date{\today}
\maketitle

\begin{abstract}
We consider the general properties of the quasispecies dynamical system from
the standpoint of its evolution and stability. Vector field analysis as well
as spectral properties of such system has been studied. Mathematical
modelling of the system under consideration has been performed.

\noindent \textbf{keywords}: quasispecies dynamics, Hamiltonian systems,
gradient dynamical system, biological evolution, complex system.
\end{abstract}

\section{Introduction}

This article is devoted to\ the theoretical study of a self-organization
problem of an ensemble of interacting species and to developing a model of a
naturally fitted coevolving ecosystem. It is well known since Eigen's work
on replicating molecules \cite{1} that the quasispecies approach is very
fruitful for modeling the fundamental behavior of evolution (See, for
example \cite{es,cam,sh,mi,1a}). Despite a huge amount of papers devoted to
this problem biological evolution is too complex that we are still far from
understanding real biological processes of self-organization. The matter is
that real experiments and obtained data on the evolution of primitive
systems need a comprehensive theoretical description that would allow to
explain these data and put them into a proper context. In this case the
central place is put to the investigation of intrinsic properties of a
nonlinear system which describe the system evolution. We put our attention
to Eigen's approach and will establish some new but very important
mathematical properties of the system which could be useful for modeling
many co-evolving ecosystem. In part we will use an approach devised in \cite%
{pg} for describing similar systems of evolution.

We start our analysis of qusispecies dynamics in molecular evolution with
discussion of background of modeling aspects which appear to be very
important for further. The first principle of modeling such types of
dynamics we accept here is based on probability consideration of molecular
replicator processes and which are well described by resulting quasispecies
concentration vector $x\in \lbrack 0,1]^{n}$, where $n\in \mathbb{Z}_{+}$ is
a number of species, being normalized by the full probability condition $%
\sum\limits_{j=1}^{n}x_{j}=1.$ The latter condition is fundamental ensuring
the full molecular number concentration and will play, in what will follow,
an important role. Moreover, the set of such vectors $x\in E^{n}$ in the
Euclidian space $E^{n}$\ forms the algebraic symplicial submanifold $%
S_{n-1}\subset E^{n}$ on which in reality the studied dynamics holds. For it
to be described analytically, it is naturally to consider a representative
symmetric matrix $P\in $End$E^{n},$ such that $P=f\otimes f,$~$<f,f>=1,$ for
same vector $f\in E^{n}$ modeling the resulting replicator dynamics
simultaneously ensuring two constraints: life on the simplex $S_{n-1}$ and
conservation of the initial molecular system information during its
replicator evolution. In general, such a dynamics can be represented in the
following form:

\begin{equation*}
P(t)=U(t)P(0)V^{-1}(t),
\end{equation*}%
with evolution parameter $t\in \mathbb{R}$, for some invertible mappings $%
V,U:\mathbb{R}\rightarrow GL(E^{n}),$ where $P(0)\in $End$E^{n}$ is an
initial molecular dynamics state. Below we consider the symmetric replicator
dynamics, whose the inverse replicator process matrix $V\in GL(E^{n})$ has
to to coincide with the forward replicator process matrix $U\in GL(E^{n}),$
that is the equality $U=V$ holds. As a result, our dynamics is representable
as
\begin{equation}
P(t)=U(t)P(0)U^{-1}(t)  \label{*1}
\end{equation}%
for all moments of time $t\in \mathbb{R}$. Assuming smooth dependence of (%
\ref{*1}) on $t\in \mathbb{R}$, one easily derives that the following Lax
type dynamics%
\begin{equation}
dP/dt=[\Lambda ,P]  \label{*2}
\end{equation}%
holds, where by definition, the matrix $\Lambda :=dU/dtU^{-1}.$ For the
matrix $P\in $End$E^{n}$ to conserve its symmetricity, the matrix $\Lambda
\in $End$E^{n}$ must be evidently skew-symmetric, that is $\Lambda =-\Lambda
^{\ast }$ in $E^{n}.$

Return now back to analyzing the intrinsic structure of our matrix $P\in $End%
$E^{n},$ modeling replicator dynamics under regard. From the general form (%
\ref{*2}) one sees that our dynamics possesses a priori so called
trace-invariants, namely, all quantities $SpP^{m}$ are such for any $m\in
\mathbb{Z}_{+}$ , where $Sp:$End$E^{n}\rightarrow \mathbb{R}$ is the
standard matrix trace operator. This fact may be naturally used within our
modeling approach. Really, consider a representative vector $f\in E^{n}$ in
the following form: $f=\{\sqrt{x_{i}}\in \mathbb{R}:i=\overline{1,n}\}.$
Then, the condition $<f,f>=$ $\sum\limits_{j=1}^{n}x_{j}=1$ is satisfied due
to \ the equality $SpP=$ $\sum\limits_{j=1}^{n}x_{j}$ and the fact that the
latter quantity is conserved! Since $P^{2}=P$, all of the other invariants
are the same, introducing into the dynamics no additional constraint. Thus,
we have naturally imbedded our molecular replicator dynamics initially \ in
the space of concentration vectors $x\in S_{n-1}\subset E^{n}$ into the
matrix phase space $\mathcal{P}\mathcal{\ni }P$ of symmetric projector
mappings of co-dimension one, that is $P=P^{\ast }$ and dim$($Im$P)=1.$ The
latter phase space $\mathcal{P}$ is still called a Grassmann manifold
possessing a lot of important intrinsic mathematical properties, which we
shell use further for deeper analyzing our molecular replicator dynamics
under regard.

Consider an evolution equation modeling our molecule replicator dynamics in
the general matrix Lax type form (\ref{*2}), where $\Lambda (x)=\{\Lambda
_{jk}:j,k=\overline{1,n}\}$ is a certain matrix\ depending on variables $%
x=\{x_{i}\in \mathbb{R}_{+}:i=\overline{1,n}\}\ $and$\ \ P=\{\sqrt{x_{j}x_{k}%
}:j,k=\overline{1,n}\}.$\ In this case as the matrix $P$\ symmetric, it is
evident that the matrix $\Lambda $\ must be skew-symmetric. Below we will
put the main attention to processes similar to qusispecies dynamics
considered in \cite{1,es}. Since we will consider $x\in \lbrack 0,1]^{n}$\
as a concentration vector of quasispecies, it has to be nonnegative for all
time, so the system (\ref{*1}) now is defined on the nonnegative orthant $%
\mathbb{R}_{+}^{n}=\left\{ x\in \mathbb{R}^{n}:~x_{i}\geq 0\right\} .$

Let us now suppose a system evolves due to the Eigen positive feedback
associated with the terms corresponding to the increase of concentrations $%
x_{j},j=\overline{1,n}:$
\begin{equation}
dx_{j}/dt=\sum\limits_{k=1}^{n}a_{jk}x_{k}R_{k}(\mathbf{x})-F_{j}(x),
\label{m2}
\end{equation}%
where an element $a_{jk}$ expresses the probability that a molecule $k$
copies into a molecule $j$ and $F_{j}(x)$ denotes the corresponding inverse
sink term. In turn the element $a_{jj}~$gives the probability that a
molecule $j$ replicates faithfully, $R_{k}(\mathbf{x})$\ is the fitness of
the molecules of $k$ type and characterizes their replication rate.

We have to determine the sink term $F_{j}(x),j=\overline{1,n}$ , in order to
fulfill the governing condition on the dynamic determined on the $(n-1)$
dimensional simplex
\begin{equation}
S_{n-1}=\left\{ x\in \mathbb{R}_{+}^{n}:\sum\limits_{i=1}^{n}x_{i}=1\right\}
.  \label{s}
\end{equation}

In order to get the corresponding source term in the equation (\ref{m2})\
let us determine elements $\Lambda _{jk},j,k=\overline{1,n}~$of the skew
symmetric matrix $\Lambda $ in (\ref{*1}) as%
\begin{equation}
\Lambda _{jk}=\frac{1}{2}\left[ a_{jk}\sqrt{x_{k}/x_{j}}R_{k}(\mathbf{x}%
)-a_{kj}\sqrt{x_{j}/x_{k}}R_{j}(\mathbf{x})\right] ,  \label{m3}
\end{equation}%
It is easy to observe here that there exists the matrix $\overline{\Lambda }%
=\{\frac{1}{2}a_{jk}\sqrt{x_{k}/x_{j}}R_{k}(\mathbf{x}):j,k=\overline{1,n}\}$%
, such that $\Lambda =\overline{\Lambda }-\overline{\Lambda }^{\ast }.$
Substituting expression (\ref{m3}) into (\ref{*1}) we get that $F_{j}(%
\mathbf{x})=x_{j}R_{j}(\mathbf{x})\sum\limits_{k=1}^{n}a_{kj},j=\overline{1,n%
}.$ As a result the governing equation for the quasispecies dynamics takes
the form:

\begin{eqnarray}
dx_{j}/dt &=&\sum\limits_{k=1}^{n}a_{jk}x_{k}R_{k}(\mathbf{x}%
)-\sum\limits_{k=1}^{n}\sqrt{x_{j}x_{k}}a_{kj}\sqrt{x_{j}/x_{k}}R_{j}(%
\mathbf{x})  \notag \\
&=&\sum\limits_{k=1}^{n}a_{jk}x_{k}R_{k}(\mathbf{x})-x_{j}R_{j}(\mathbf{x}%
)\sum\limits_{k=1}^{n}a_{kj}  \notag \\
&=&\sum\limits_{k\neq j}^{n}(a_{jk}x_{k}R_{k}(\mathbf{x})-x_{j}R_{j}(\mathbf{%
x}))a_{kj}  \label{m4}
\end{eqnarray}%
It is easy to see after summing up equations (\ref{m4}) that

\begin{equation*}
\sum\limits_{j}^{n}dx_{j}/dt=0,
\end{equation*}%
meaning in this case that the evolution dynamics is determined really on the
simplex (\ref{s}).

It should be mentioned that the system (\ref{m4}) is a little bit diverse
from systems considered in a set of papers \cite{es,ems,cam,sh,mi,1a} mainly
by sink terms. But namely only such a form of the sink term ensures the
important simplex condition $\sum\limits_{j}^{n}x_{j}=1$ for all $t\in
\mathbb{R}$, without additional constraints involved in the model.

Note here that, the system (\ref{m4}) is really representable in the
evolution form (\ref{*1}) that can be checked easily, and the matrix $P\in $%
End$\mathbb{R}^{n}$ is a one dimensional symmetric projector, that is $%
P^{2}=P,~P^{\ast }=P$ for $t\in \mathbb{R}$, being important for our further
studying the structure of vector field (\ref{*1}) on the corresponding
projector matrix manifold $\mathcal{P}$ \cite{5,7}. The component vector
form of the system (\ref{m4}) can be also represented as%
\begin{equation}
d\mathbf{x}/dt=AR(\mathbf{x})\mathbf{x-}BR\mathbf{\mathbf{(\mathbf{x})}x.}
\label{m5}
\end{equation}%
with $A=\{a_{jk}:j,k=\overline{1,n}\},$ $B=diag\left\{ \underset{k=1}{%
\overset{n}{\sum }}a_{kj}:j=\overline{1,n}\right\} ,$ and~$R(\mathbf{x}),~%
\mathbf{x\in }S_{n-1}$ being a fitness matrix \ expression. If the fitness
matrix $R(\mathbf{x}),$\ $\mathbf{x\in }S_{n-1}$\ is diagonal one that is $R=%
\bar{R}=diag\left\{ \bar{R}_{j}:j=\overline{1,n}\right\} $\ and does not
depend on $\mathbf{x\in }S_{n-1},$ the system (\ref{m5}) evidently will be
linear. Thereby the solution of such a system can be obviously represented
as
\begin{equation*}
\mathbf{x(}t\mathbf{)=x}(0)~exp[(A-B)\bar{R}]t,
\end{equation*}%
where $x(0)\mathbf{\in }S_{n-1}$ is a concentration of population of each
type at initial time.

\section{Vector field analysis: imbedding into gradient structure}

Since our system dynamic flow (\ref{*2}) in reality lives on the projector
matrix of Grassmann manifold $\mathcal{P}\mathcal{\ni }P$ \cite{8}, all its
properties can be naturally extracted from deeper analysis of its structure
on this manifold. In particular, it is important to know where the vector
field (\ref{m4}) admits the structure of gradient type endowed \ with some
Lyapunov function, ensuring the existence of stable states on the compact
Grassmann manifold $\mathcal{P}$. In order to study the gradient field
structure of the flow (\ref{*1}) if any on the projector matrix manifold $%
\mathcal{P}\mathcal{\ni }P$ let us consider a smooth functional $\Psi :%
\mathcal{P\rightarrow }\mathbb{R},$ whose usual variation is given as

\begin{equation}
\delta \Psi (P):=\text{Sp}(D\delta P)  \label{p7}
\end{equation}%
with a symmetric matrix $D\in $End$E^{n}$ and Sp$:$End$E^{n}\mathcal{%
\rightarrow }\mathbb{R}^{1}$ being as before the standard matrix trace.
Taking into account the natural metrics on $\mathcal{P}$, \ we \ consider
the projection $\nabla _{\varphi }\Psi $\ of the usual gradient vector field
$\nabla \Psi $ upon the tangent space $T(\mathcal{P})$ under the following
conditions:

\begin{equation}
\varphi (X;P):=\text{Sp}(P^{2}-P,X)=0,\ \text{Sp}(\nabla \varphi ,\nabla
_{\varphi }\Psi )\mid _{\mathcal{P}}=0,  \label{p8}
\end{equation}%
holding on $\mathcal{P}$ for all $X\in $End$\mathbb{R}^{n}$.$\;$The first
condition is evidently equivalent to $P^{2}-P=0,$ that is $P\in \mathcal{P}$%
. Thereby we can formulate such a Statement.

\textbf{Statement 1}. The functional gradient $\nabla _{\varphi }\Psi (P),$ $%
P\in \mathcal{P}$ under the condition (\ref{p8}) admits the following
commutator Lax type representation:%
\begin{equation*}
\nabla _{\varphi }\Psi (P)=\left[ \Lambda ,P\right]
\end{equation*}%
with $\Lambda \in $end$E^{n}$\ being a skew-symmetric matrix satisfying the
commutator equation
\begin{equation*}
\Lambda =[D,P],
\end{equation*}
where $D$\ is a symmetric matrix. Really, consider the projection of the
usual gradient $\nabla \Psi (P)$ upon the tangent space $T(\mathcal{P})$ of
the Grassmann manifold $\mathcal{P~}$with $P\in \mathcal{P}$ imbedded
into~End$E^{n}$:

\begin{equation}
\nabla _{\varphi }\Psi (P)=\nabla \Psi (P)-\nabla \varphi (Q;P),  \label{p10}
\end{equation}%
where $Q\in $End$E^{n}$ is some still unknown matrix. Taking into account
the conditions (\ref{p8}), we find that

\begin{eqnarray}
\nabla _{\varphi }\Psi (P) &=&D-Q-P(D-Q)-(D-Q)P+PD+DP  \notag \\
&=&PD+DP+2PQP,  \label{p12a}
\end{eqnarray}%
where\ we made use of the relationships%
\begin{equation*}
\nabla _{\varphi }\Psi (P)=D-Q+PQ+QP)
\end{equation*}%
and
\begin{equation*}
P(D-Q)+(D-Q)P+2PQP=D-Q.
\end{equation*}%
Now one can easily see from (\ref{p12a}) and the second condition in (\ref%
{p10}), that

\begin{equation}
PQP=-PDP  \label{p13}
\end{equation}%
for all $P\in \mathcal{P}$, giving rise to the final result

\begin{equation}
\nabla _{\varphi }\Psi (P)=PD+DP-2PDP,  \label{p14}
\end{equation}%
coinciding exactly with the commutator $[[D,P],P].$ Since the matrix admits
the representation
\begin{equation*}
\Lambda =\bar{\Lambda}-\bar{\Lambda}^{\ast }=DP-PD
\end{equation*}%
one gets right away that
\begin{equation}
\bar{\Lambda}=DP+S  \label{lam}
\end{equation}%
with $S=S^{\ast }\in $End$E^{n}.$\ If there exists such symmetric
matrices $D$ and $S\in $End$E^{n}$ \ that (\ref{lam}) is
satisfied, then our model will be evidently of gradient type. In
particular, the matrix $D\in $End$E^{n}$ found from (\ref{lam})
must enjoy the Volterra criteria $D^{\prime }(P)=D^{\prime \ast
}(P)$ for any $P\in \mathcal{P}$ imbedded into End$E^{n}. $

It should be noted here that the Grassmann manifold $\mathcal{P}$ is also a
symplectic manifold \cite{5,7} whose canonical symplectic structure is given
by the expression:

\begin{equation}
\omega ^{(2)}(P):=\text{Sp}(PdP\wedge dPP),  \label{p15}
\end{equation}%
where $d\omega ^{(2)}(P)=0$ \ for all $P\in \mathcal{P}$, and the
2-differential form (\ref{p15}) is non-degenerate \cite{5,8} upon the
tangent space $T(\mathcal{P}).$

Let us assume now that $\xi :\mathcal{P\rightarrow }\mathbb{R}$ is an
arbitrary smooth function on $\mathcal{P}$. Then the Hamiltonian vector
field $X_{\xi }:\mathcal{P\rightarrow }T(\mathcal{P})$ on $\mathcal{P}$
generated by this function subject to the symplectic structure (\ref{p15})
is given as follows:

\begin{equation}
X_{\xi }=[[D_{\xi },P],P],  \label{p16}
\end{equation}%
where $D_{\xi }\in $End$E^{n}$ is a certain symmetric matrix. The vector
field $X_{\xi }:\mathcal{P\rightarrow }T(\mathcal{P})$ generates on the
compact manifold $\mathcal{P}$ the flow

\begin{equation}
dP/dt=X_{\xi }(P),  \label{p17}
\end{equation}%
being defined globally for all $t\in \mathcal{\ }\mathbb{R}$. This flow by
construction is evidently compatible with the projector condition $P^{2}=P$.
This means in particular that the condition

\begin{equation}
-X_{\xi }+PX_{\xi }+X_{\xi }P=0  \label{p18}
\end{equation}%
holds on $\mathcal{P}$. Thus, we stated that dynamical system (\ref{*2})
being considered on the Grassmann manifold $\mathcal{P~}$can be Hamiltonian
that makes it possible to formulate the following statement.

\textbf{Statement 2.} A gradient vector field of the form (\ref{p16}) on the
Grassmann manifold \ $\mathcal{P}$ is Hamiltonian with respect to the
canonical symplectic structure (\ref{p15}) and a certain Hamiltonian
function $\xi :\mathcal{P\subset }$end$E^{n}\mathcal{\rightarrow }\mathbb{R}$%
, satisfying conditions

\begin{eqnarray*}
\nabla \xi (P) &=&[D_{\xi },P]-Z+PZ+ZP, \\
\ \nabla \xi ^{\prime }(P) &=&\nabla \xi ^{\prime \ast }(P),\ D_{\xi }=D
\end{eqnarray*}%
for some matrix $Z\in $End$E^{n}$\ for all $P\in \mathcal{P\subset }E^{n}$.

Consider now the $(n-1)$-dimensional simplex $S_{n-1}$ as a Riemannian space
$M_{g}^{n-1}=S_{n-1}$with the metrics

\begin{equation*}
ds^{2}(x):=d^{2}\Psi \mid _{\mathcal{P}}(x)=\underset{i,j=1}{\overset{n}{%
\sum }}g_{ij}(x)dx_{i}dx_{j}\mid _{\mathcal{P}},
\end{equation*}%
where for $i,j=\overline{1,n},$ $g_{ij}(x)=\frac{\partial ^{2}\Psi (x)}{%
\partial x_{i}\partial x_{j}},\ \ \underset{i=1}{\overset{n}{\sum }}x_{i}=1.$

Relative to the metrics on $M_{g}^{n-1}$ we can calculate the gradient $%
\nabla _{g}\Psi $ of the function $\Psi :\mathcal{P\rightarrow }\mathbb{R}$
and set on $M_{g}^{n-1}$ the gradient vector field

\begin{equation}
dx/dt=\nabla _{g}\Psi (x),  \label{p19}
\end{equation}%
\ with $x\mathcal{\in }M_{g}^{n-1}$, that is the condition $\underset{i=1}{%
\overset{n}{\sum }}x_{i}=1$ is satisfied a priory. Having calculated (\ref%
{p19}), we can formulate the next statement.

\textbf{Statement 3.} The gradient vector fields $\nabla _{\varphi }\Psi $
on $\mathcal{P}$ and $\nabla _{g}\Psi $ on $M_{g}^{n-1}$ are equivalent, or
in another words, vector fields

\begin{equation}
dx/dt=\nabla _{g}\Psi (x)  \label{p20}
\end{equation}%
and
\begin{equation}
dP(x)/dt=\left[ \left[ D(x),P(x)\right] ,P(x)\right]  \label{p20a}
\end{equation}%
\ generate the same flow on $M_{g}^{n-1}$.

As a result from the Hamiltonian property of the vector field $\nabla
_{\varphi }\Psi $ on the Grassmann manifold $\mathcal{P}$ we get such a new
statement.

\textbf{Statement 4.} The gradient vector field $\nabla _{g}\Psi $ (\ref{p19}%
) on the metric space $M_{g}^{n-1}$ where $n=2m+1$ is Hamiltonian subject to
the non-degenerate symplectic structure

\begin{equation}
\omega _{g}^{(2)}(x):=\omega ^{(2)}(P)\mid _{M_{g}^{n-1}}  \label{p21}
\end{equation}%
for all $x\mathcal{\in }M_{g}^{2m}$ with a Hamiltonian function $\xi _{\psi
}:M_{g}^{2m}\mathcal{\rightarrow \mathbb{R}}$, where $\xi _{\psi }:=\xi \mid
_{M_{g}^{n-1}},$ $\xi :\mathcal{P\rightarrow }\mathbb{R}$ is the Hamiltonian
function of the vector field $X_{\xi }$ (\ref{p16}) on $\mathcal{P}$.
Otherwise, if $n\in \mathbb{Z}_{+}$ is arbitrary \ our two flows (\ref{p20})
and (\ref{p20a}) are on $\mathcal{P}$ only Poissonian.

\section{Spectral properties}

Consider the eigenvalue problem for a matrix $P\in \mathcal{P}$, depending
on the evolution parameter $t\in \mathbb{R}$:

\begin{equation}
P(t)f=\lambda f,  \label{p22}
\end{equation}%
where $f\in \mathbb{R}^{n}$ is an eigenfunction, $\lambda \in \mathbb{R}$ is
a real eigenvalue since $P^{\ast }=P$, i.e. matrix $P\in \mathcal{P}$ is
symmetric. It is seen from the expression $P^{2}=P$ that spec$P(t)=\left\{
0,1\right\} $ for all $t\in \mathbb{R}$. Moreover, taking into account the
invariance of Sp$P=1$ we can conclude that \ only one eigenvalue of the
projector matrix $P(t),$ $t\in \mathbb{R}$, is equal to 1, all others being
equal to zero. \

In general case the image Im$P\subset \mathbb{R}^{n}$ of the matrix $P(t)\in
\mathcal{P}$ for all $t\in \mathcal{\ }\mathbb{R}$ is $k-$dimensional, $%
k=rankP,\ $and the kernel Ker$P\subset \mathbb{R}^{n}$ is $(n-k)$%
-dimensional, where $k\in \mathbb{Z}_{+}$ is constant, not depending on $%
t\in \mathcal{\ }\mathbb{R}$. As a consequence we establish that at $k=1$
there exists a unique vector $f_{0}\in \mathbb{R}^{n}/($Ker$P)$ for which

\begin{equation}
Pf_{0}=f_{0},\   \label{p23}
\end{equation}%
Due to the statement above for a projector $P\in $End$E^{n}$ we can write
down the following expansion in the direct sum of mutually orthogonal
subspaces: $E^{n}=$Ker$P\oplus $Im$P.$

Take now $f_{0}\in \mathbb{R}^{n}$ satisfying the condition (\ref{p23}).
Then in accordance with (\ref{p20}) the next statement holds.

\textbf{Statement 5.} The vector $f_{0}\in E^{n}$ satisfies the following
evolution equation:

\begin{equation}
df_{0}/dt=\left[ D(x),P(x)\right] f_{0}+C_{0}(t)f_{0},  \label{p25}
\end{equation}%
where $C_{0}:\mathbb{R}\mathcal{\rightarrow }\mathbb{R}$ is a certain
function depending on the choice of the vector $f_{0}\in $Im$P$. At some
value of the vector $f_{0}\in $Im$P$ we can evidently ensure the condition $%
C_{0}\equiv 0$ for all $t\in \mathcal{\ }\mathbb{R}$. Moreover one easily
observes that for the matrix $P(t)\in \mathcal{P}$ \ one has \cite{7} \ the
representation $P(t)=f_{0}\otimes f_{0}$, $\left\langle
f_{0},f_{0}\right\rangle =1$, giving rise to the system (\ref{m4}) if $%
f_{0}:=\left\{ \sqrt{x_{j}}\in \mathbb{R}_{+}:j=\overline{1,n}\right\} \in
E^{n}$.

\section{Discussion of the quasispecies dynamics}

Quasispecies dynamics is very interesting for researchers and there are a
lot of papers devoted to computer simulation of it \cite{es,cam,sh,mi,1a}.
In order to single out the characteristic features of the model stated above
let us write down the model typically used for quasispecies dynamics. In
vector form such a model can be written as

\begin{equation}
d\mathbf{x}/dt=(AR-<R>)\mathbf{x}  \label{si1}
\end{equation}%
where $R=diag\left\{ \bar{R}_{j}:j=\overline{1,n}\right\} $\ is the diagonal
matrix with the Malthusian fitness values, $\ <R>:=\sum_{i=1}^{n}\bar{R}%
_{i}x_{i}.$ The diagonal elements $a_{jj},$ $j=\overline{1,n},$ of the
matrix $A=\{a_{jk}:j,k=\overline{1,n}\}.$ In correspond to the
self-replication process and nondiagonal -\ to mutation. In order to fulfill
the simplex condition we have to put \cite{es,cam,sh}\ for each column $%
a_{jj}=1-\sum\limits_{k=1}^{n}a_{kj},~j=\overline{1,n}$ and hence the column
sum will give rise to unity. In contrast to (\ref{si1}) in our model (\ref%
{m4}) we do not require such a constraint in order to satisfy the simplex
condition.

The mutation matrix $A$ in our model has to be nondiagonal. The choice of
matrix $A$ specifies the generation and recombination rates among different
molecules in chemistry. A very interesting application of the model stated
in the papers \cite{es,cam,sh}\ is for describing quasispecies evolution. In
framework of this approach the bio molecules are considered as a bitstrings
of length $L$. In this case we have $2^{L}$ different molecules. As a result
the mutation matrix $A$ is of huge dimension. If we consider a bio molecules
with $L>>1$ this approach is practically out of reason and can not be
treated for large bio macromolecule. In this case some simplification has
been considered when certain macromolecules are grouped together in such a
way that the number of independent variables is reduced to $L+1$ \cite{ss}.
In the framework of this approach a certain sequence (master sequence) is
chosen beforehand and other sequences are grouped all into error classes,
according to their Hamming distance from the chosen one. Sequences which
have the same Hamming distance from the master one compile a one error
class. Such reduction of the dimensionality make it possible to get the
problem acceptable for computer simulation and to reveal some features
inherent to real bio world. In this case we can write down the nondiagonal
elements $a_{jk}$\ of mutation matrix $A$ as the probability of mutation
string $k$ to string $j$
\begin{equation}
a_{jk}=q^{L-H_{jk}}(1-q)^{H_{jk}}  \label{ss3}
\end{equation}%
\begin{figure}[tbp]
\begin{center}
\begin{tabular}{c}
\begin{minipage}[t]{0.5\textwidth} \centerline{\epsfxsize=0.9\textwidth
\epsfbox{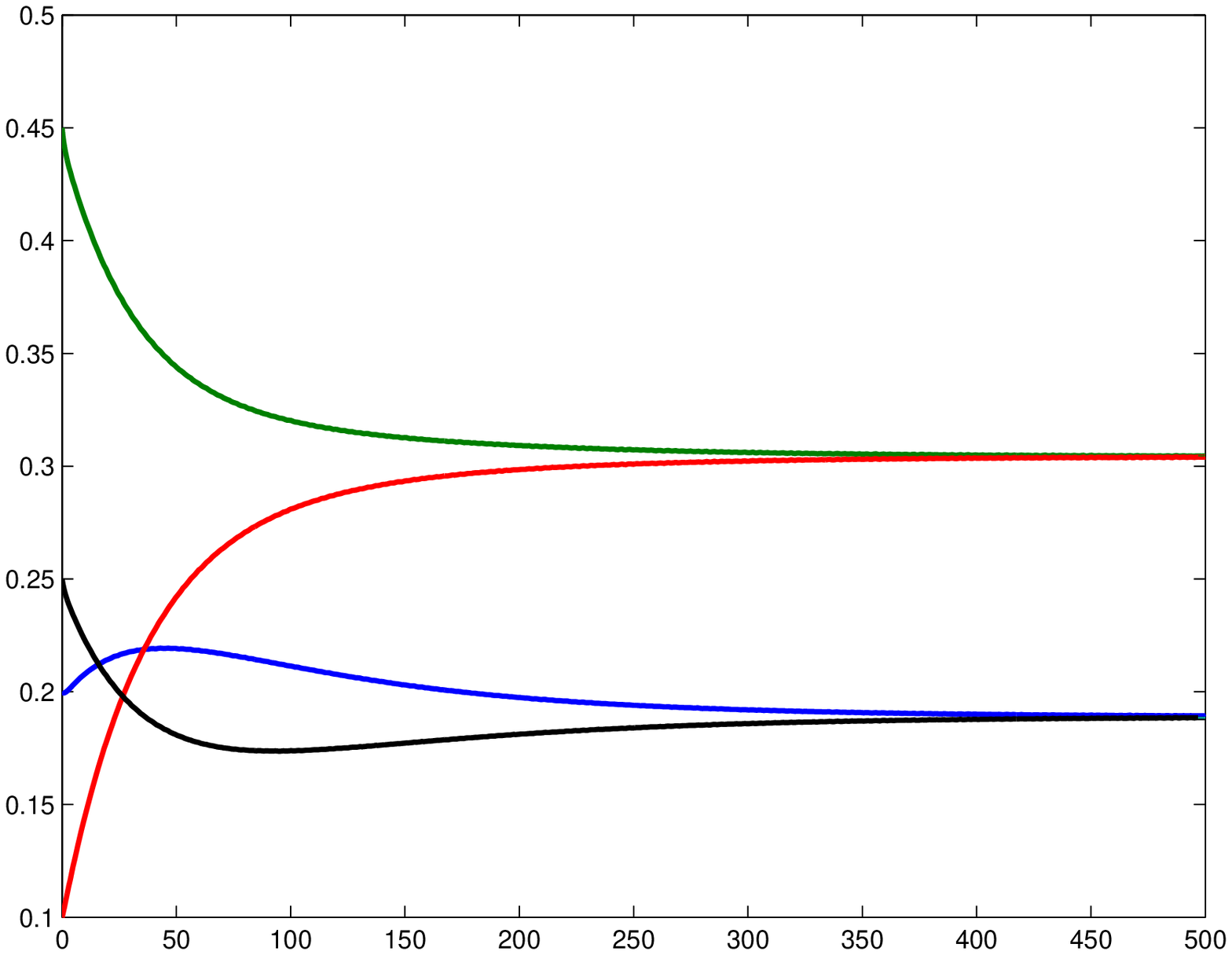 a}} \end{minipage} \begin{minipage}[t]{0.5\textwidth}
\centerline{\epsfxsize=0.9\textwidth \epsfbox{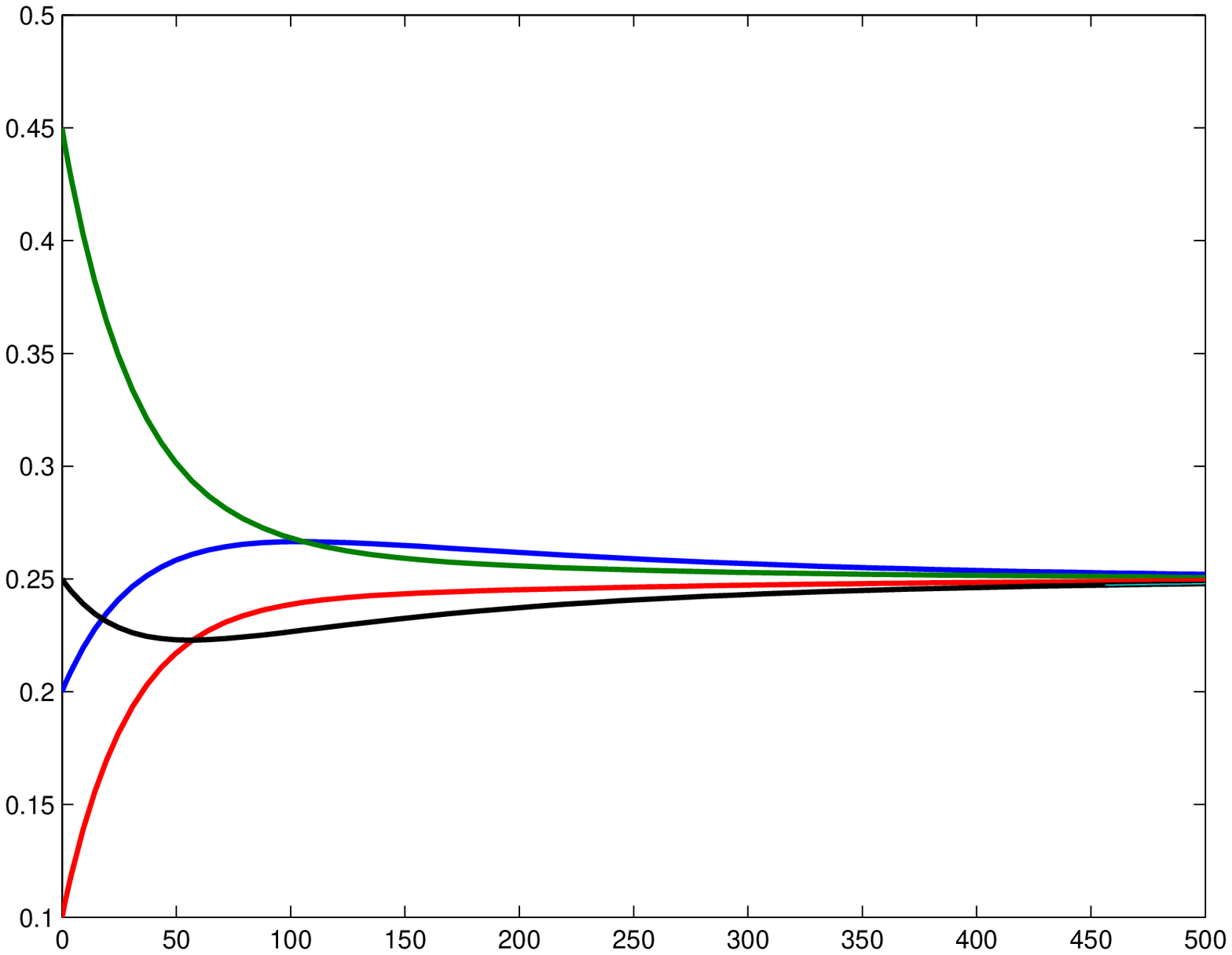 b}} \end{minipage}%
\end{tabular}%
\end{center}
\caption{Quasispecies dynamics as a results of computer simulation of the
equation (\protect\ref{si1})- a and our model (\protect\ref{m4}) - b.
Initial conditions: $\mathbf{x}=(0.2;0.45;0.1;0.25)^{T}$}
\end{figure}
where $q$ is the probability that a particular locus of the chain is copied
correctly into the next generation i.e. the probability of replication, ($%
1-q)$ is the probability of mutation), $L$ is a bit string length $H_{jk}$\
is Hamming distance between string $j$ and $k$ and is defined as the number
of positions, in which the two sequences differ. In order to note the
difference between two model we simulate our model for simplest case $L=3.$
The plot of computer simulation with matrix given by formulae (\ref{ss3}) is
presented on Fig.1 (a, b). We can easily see that having the same initial
conditions and matrix replication mutation matrix $A$ the results of the
evolution process is quite different.

Let us find a stationary frequency distribution in the framework of derived
model (\ref{m4}). If we make set the right hand side of the equation (\ref%
{m4}) to zero we get for constant fitnesses the next expression

\begin{equation}
x_{j}=\frac{\sum\limits_{k\neq j}^{n}a_{jk}x_{k}\bar{R}_{k}}{\bar{R}%
_{j}\sum\limits_{k\neq j}^{n}a_{kj}}.  \label{si2}
\end{equation}%
Analyzing the expression (\ref{si2}) one can see that the values of the
frequencies do not depend on diagonal elements $a_{jj},~~j=\overline{1,n},$
and replicator process in this model is determined only by the input given
by nondiagonal elements $a_{kj},~k\neq j=\overline{1,n},$ being natural from
the biological point of \ view. If the fitness value of some species $\bar{R}%
_{j}$\ is much greater than all of the rest fittnesses, the values of $x_{j}$%
\ will be chosen less than all the others. In this case the results obtained
for example for a single peak model landscape in the framework of (\ref{si1}%
), which lead to so called \ \textquotedblleft phase
transition\textquotedblright\ and vanishing of the corresponding species
named there master ones if $a_{jj}<1/\bar{R}_{j}$ be interpreted in another
way \cite{es,cam,sh}$.$

The equation (\ref{si2}) can be evidently written down as a fixed point
problem $x=\tilde{A}x,$ where $\tilde{A}=\{a_{jk}\bar{R}_{k}/(\bar{R}%
_{j}\sum\limits_{k\neq j}^{n}a_{kj}):k,j=\overline{1,n}\}$ with diagonal
elements $a_{jj}:=0,$ $j=\overline{1,n}.$ So, its solution exists if the
matrix $\tilde{A}$ possesses the eigenvalue $\lambda =1,$ or the determinant
equation $det(\mathbf{1}-\tilde{A})=0,$ where $\mathbf{1}$ is unity matrix,
is satisfied identically. But this fact is true for any matrix $A$ and
arbitrary parameters $\bar{R}_{j}\in \mathbb{R}_{+},j=\overline{1,n}.\ $If
this equation is not satisfied, the process of reducing some amount of
species from the system happens, that is some of frequencies will become
exactly zero \ and the resulting system remains to live on a simplex of
lower dimension. The latter situation can be considered as a
\textquotedblleft phase reduction\textquotedblright\ \ naturally related
with some threshold values of frequencies found from the determinant
equation written above, against the notion of \textquotedblleft phase
transition\textquotedblright\ \ used in cited above articles. This behavior
is of great interest for diverse applications since it can be interpreted as
some kind of simplex reduction $S_{n-1}\rightarrow S_{n-l}$ for $1<l<n,$
taking place in some kinds of replicator dynamics models. Taking into
account this aspect of these models and their virtual importance in studying
biological replicator and other models, we plan to dwell on their deeper
studying soon in another place.

\textbf{Acknowledgements.} Authors are kindly indebted to their friends, in
particular prof. Z. Peradzhynski from Warsaw University, professors J.
Myjak, S. Brzychczy and B. Choczewski from AGH University of Science and
Technology of Krakow for \ fruitful and constructive discussions of the
replicator dynamics models considered in this paper. The work of A.P. was in
part financed through a local WMS\_AGH grant for which he cordially thanks.

\end{document}